\documentclass[a4paper]{jpconf}
\usepackage{graphicx}
\usepackage{amsmath}
\usepackage{subfigure}

\newcommand{\naitl}{NaI(Tl)}
\newcommand{\mili}{$10^{-3}$}
\newcommand{\tmili}{$\times 10^{-3}$}
\newcommand{\kevee}{keV$_\text{ee}$}

\begin{document}
\title{SABRE: WIMP modulation detection in the northern and southern hemisphere}

\author{F. Froborg, for the SABRE Collaboration}

\address{Department of Physics, Princeton University, Princeton, NJ 08544, USA.}

\ead{ffroborg@princeton.edu}

\begin{abstract}
Measuring an annual modulation in a direct Dark Matter detection experiment is not only a proof of the existence of WIMPs but can also tell us more about their interaction with standard matter and maybe even their density and velocity in the halo. Such a modulation has been measured by the DAMA/LIBRA experiment in \naitl\ crystals. However, the interpretation as WIMP signal is controversial due to contradicting results by other experiments. 
The SABRE experiment aims to shed light on this controversy by detecting the annual modulation in the same target material as DAMA with twin detectors at LNGS in Italy and at SUPL in Australia.
The two locations in the northern and southern hemisphere allow to verify if other seasonal effects or the site have an influence on the measurement, thus reducing systematic effects. This paper will give an overview on the experimental design, the current status of the proof of principle phase mainly devoted to high-purity crystal growing, and an outlook on future plans.
\end{abstract}

\section{Motivation}
For many years now dark matter experiments have searched for nuclear recoils produced by the leading dark matter candidate, the Weakly Interacting Massive Particle (WIMP), due to the motion of the Earth through the hypothetical WIMP halo. One of these experiments is called DAMA (short for DAMA/NaI and DAMA/LIBRA), which has been measuring a modulation in their signal for over a decade now, using NaI(Tl) crystals \cite{dama2013}. However, the interpretation as a WIMP signal is controversial due to contradictory results by other experiments such as LUX, XENON, or SuperCDMS \cite{lux, xenon100, supercdms}. These experiments, though, rely on different target materials such that a direct comparison of the results is model dependent. A measurement with the same target material is therefore essential to better understand the origin of the modulation. Such a measurement is in preparation with the SABRE experiment \cite{sabre2015}. If the modulation is indeed due to WIMPs, a precision measurement might tell us more about their interaction with standard matter and maybe even their density and velocity in the galactic halo. It might also give us further insight into the proposed focusing effect of the WIMP wind due to the gravitational potential of the sun \cite{Gluscevic2015, Lee2013}. Other, new physics might be involved, which is equally an exciting prospect - after all, few doubt that DAMA do indeed measure an annual modulation - and no-one has so far proposed a convincing explanation for this effect.

\section{The SABRE Experiment}
To be able to test DAMA's results, the Sodium Iodide with Active Background REjection (SABRE) experiment plans to improve DAMA's setup by concentrating on three main focus areas: higher purity crystals, better performance, and lower background. These focus areas are described in the following, and it is also shown what can be expected with this setup.

\subsection*{High Purity Crystals}
A crucial part of this effort is the development of high-purity crystals since DAMA's crystal grower  is not selling their high-purity crystals to anyone else. Thus, we conducted our own R\&D, which started with the development of ultra-high purity NaI powder. As a result of this effort, our industrial partner Sigma-Aldrich managed to produced the so-called Astrograde powder with K levels as low as 3.5~ppb. Table~\ref{tab:powder} shows inductively coupled plasma mass spectroscopy (ICP-MS) results (confirmed by gamma counting) for the most interesting impurities K, Rb, U, and Th for the Astrograde powder in comparison with DAMA's powder and crystal purity. The values show that the Astrograde powder has significantly lower impurity levels compared to DAMA's powder and similar levels as DAMA's crystals. It also shows that DAMA's crystal grower Saint Gobain managed to reduce the impurity levels significantly during the growth process, indicating that crystals grown from Astrograde powder can result in crystals with lower impurities than DAMA's. We are currently working on methods to further purify the NaI prior to crystal growth to further reduce the impurities.
\begin{table}[!b]
\centering
\caption{Purity levels of Sigma Aldrich's Astrograde NaI powder in comparison with DAMA's powder and crystal purity \cite{damaapp}.}
\label{tab:powder}
\begin{tabular}{c cccc}
\br
			    & Sigma-			 & DAMA  		  & DAMA \\
Element 	& Aldrich [ppb] & Powder [ppb] & Crystal [ppb]\\\mr
K 				& 3.5 (18)* & 100 & $\sim$13\\
Rb				& 0.2 & n.a. & $< 0.35$\\
U 				& $< 1.7$ ($<$ \mili)** & $\sim 0.02$ & 0.5 --7.5\tmili \\
Th 			& $<$ 0.5 ($<$ \mili)** & $\sim 0.02$ & 0.7 --10\tmili \\\br
\end{tabular}\\*[2mm]
{\footnotesize * Independent measurement\quad ** Preliminary measurement at PNNL; full validation needed.}
\end{table}

As the next step we are developing together with Radiation Monitoring Devices (RMD) a crystal growth procedure, which at least does not introduce any impurities and preferably even reduces the impurity levels similar to DAMA's procedure. We are using the vertical Bridgman-Stockbarger technique \cite{Bridgman1925} to grow our crystals, where a crucible is typically placed inside a sealed ampoule. The sealed environment has the advantage that it reduces the possibility of contaminate the material during the growth phase. To determine the best crucible-ampoule combination, several materials were prepared with different cleaning procedures. Then, small test growths using Astrograde powder were done to determine the radioactive contamination of the crystal during the growth via ICP-MS. The optimum crucible-ampoule combination together with a precision cleaning showed no increase of the impurity levels inside the crystal. The result was confirmed by a blank test, which includes all steps of the crystal growth but uses no powder and is followed by a more sensitive leach test. The small test growth also allowed us to evaluate the crystal growing procedure and optimize it in terms of possible contamination.

The found procedure is now in the process of being scaled up to larger multi-kilogramm crystals. First growths were performed using standard purity powder and a crystal was grown with $\sim8.5$~cm diameter and $\sim10$~cm length ($\sim2$~kg). We are currently in the process of growing a high-purity crystal with the evaluated procedure and results can be expected soon. The final goal is to grow $\sim5$-kg high-purity crystals.

\subsection*{Improved Performance}
To improve the overall performance of the experiment compared with DAMA, SABRE plans to directly couple two 3" photomultiplier tubes (PMTs) to the crystal as shown in Figure~\ref{fig:CTFsetup} on the left. The PMTs will have a high quantum efficiency ($\sim35$\%) combined with the lowest possible background. Our collaboration with Hamamatsu Japan has already resulted in PMTs with a low radioactivity of $\sim 3$~mBq/PMT U, $\sim0.5$~mBq/PMT Th, Co and $\sim 2$~mBq/PMT K \cite{pmtback_2015}, the R11065-20 series. Additional improvements are currently in development, which are expected to further reduce the total radioactivity by a factor of $\sim10$ and might also improve the overall stability of the PMTs.

To reduce afterglow in the PMTs, we plan on running them on a significantly lower voltage where the afterglow noise is at a lower level than random coincidences of dark counts. This is possible due to cold preamplifiers developed by our collaborators at LNGS, Italy.

\subsection*{Lower Background}
The most dangerous isotope in terms of background is $^{40}$K since one of its decay channels emits 3~keV X-ray/Auger electrons, which produce events right in the region of interest at 2--6~keV. The 3~keV electron is accompanied by a 1.46~MeV gamma, which can be used to identify and ultimately veto these events. DAMA partially removes these events by only allowing single crystal events to contribute to their signal. However, the outer crystals in particular are not well covered, leaving a significant background contribution in the region of interest. 
\begin{figure}[!t]
	\begin{minipage}{.68\textwidth}
		\includegraphics[height=.23\textheight]{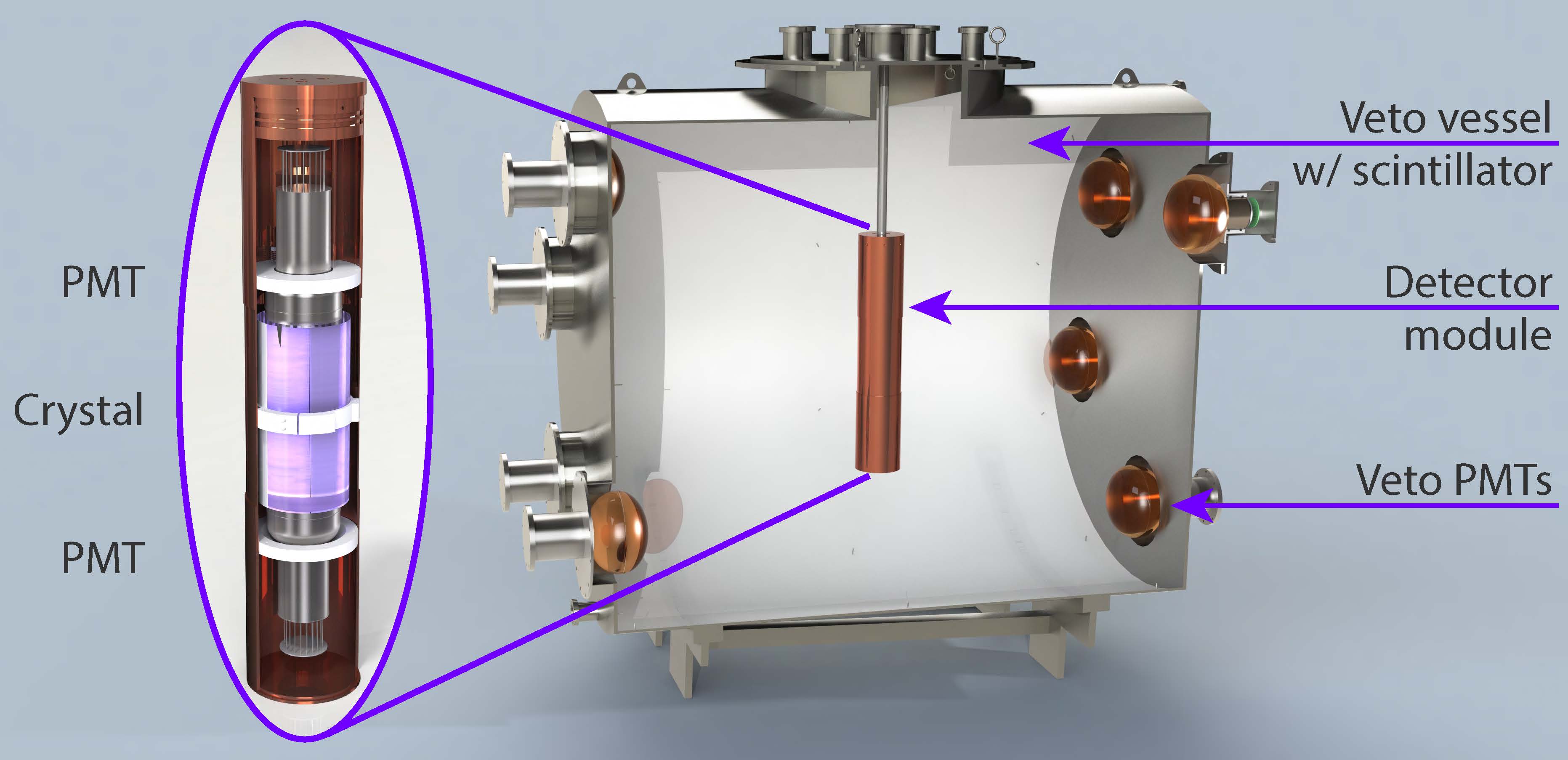}	
	\end{minipage}
	\begin{minipage}{.1\textwidth}
		\includegraphics[height=.23\textheight]{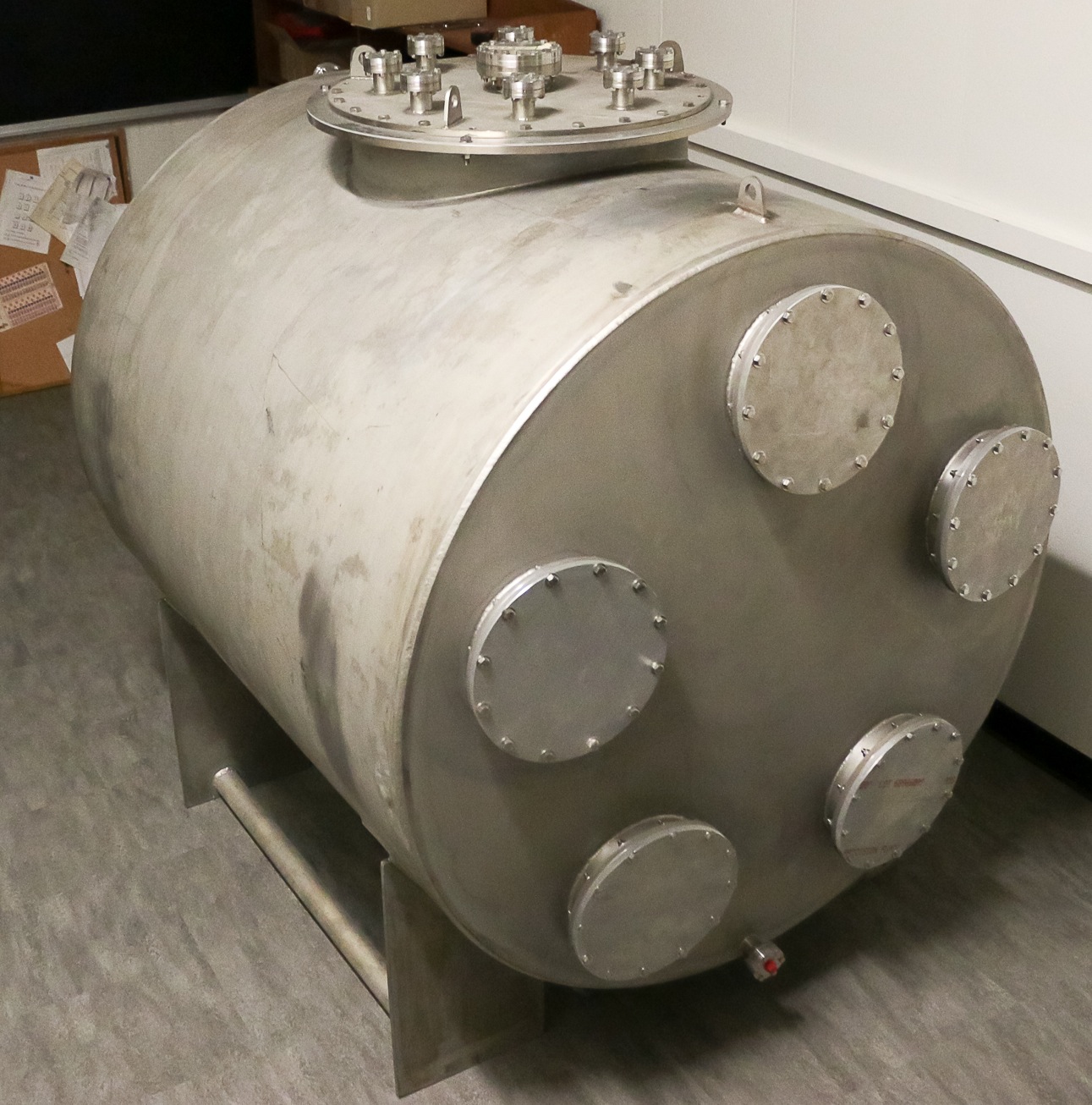}
	\end{minipage}
	\caption{Setup of SABRE: A crystal is directly coupled to 2 PMTs and placed inside a copper enclosure (shown on the left). This detector module will be placed inside the liquid scintillator veto, a stainless steel vessel equipped with PMTs and filled with pseudocumene (photograph on the right, assembly in the middle).}
	\label{fig:CTFsetup}
\end{figure}

SABRE plans to reduce such background contributions by submerging the detector module(s) into a $1.4\text{ m dia.} \times 1.5$~m cylindrical liquid scintillator veto as shown in Figure~\ref{fig:CTFsetup}. The liquid scintillator Veto (right) is a low-radioactivity stainless steel tank containing $\sim$2 tons of liquid scintillator (pseudocumene with PPO) viewed by 10 Hamamatusu R5912 PMTs.  The vessel's inner surface is coated with ETFE  to prevent scintillator degradation due to contact with the stainless steel.  The predicted light yield of the liquid scintillator veto is 0.22~photoelectrons/\kevee\ with a Lumirror liner on the inner surface\cite{EmilyThesis}. The veto vessel will be further shielded from external radiation by a combination of lead, water, and polyethylene.

\subsection*{Expectations}
\begin{figure}[t]
	\centering
	\subfigure[Background estimate.]{
		\includegraphics[height=.22\textheight]{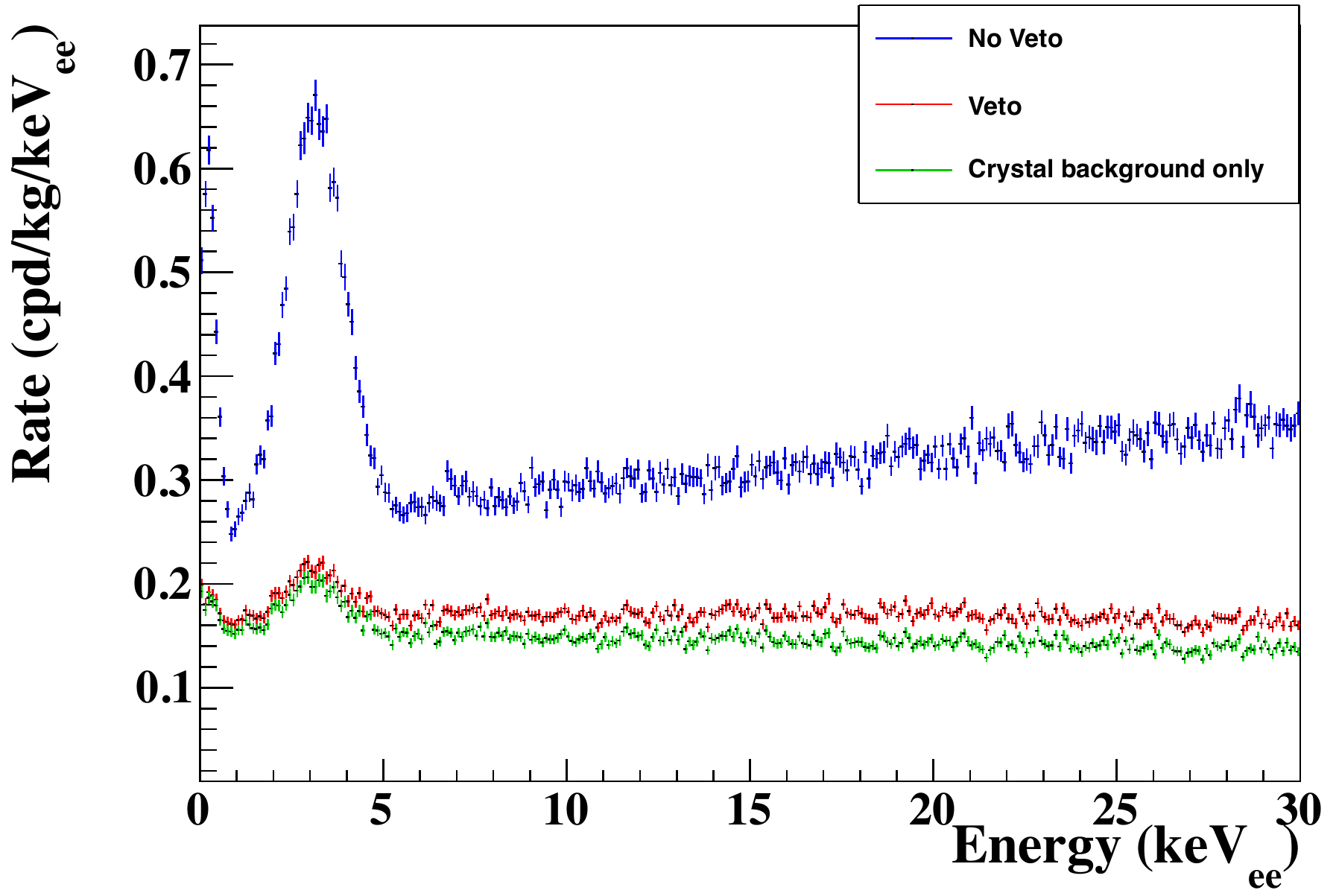}
		\label{fig:est-back}}
	\hfill
	\subfigure[Estimated sensitivity.]{
		\includegraphics[height=.22\textheight]{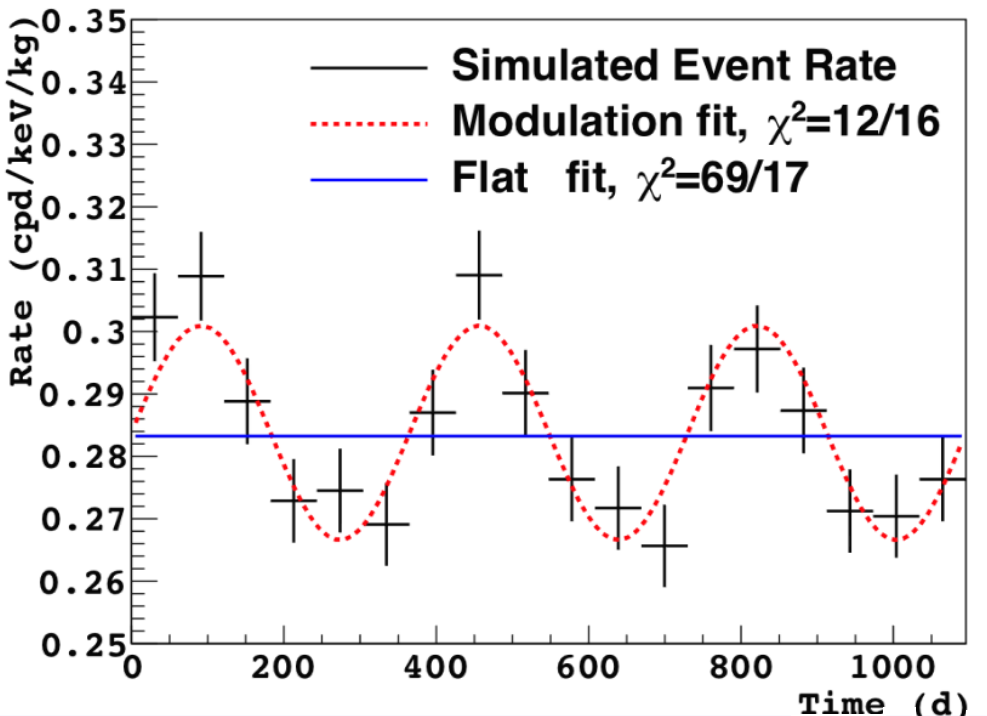}
		\label{fig:est-sens}}
	\caption{Estimated background and sensitivity for SABRE with 50-kg target material based on Geant4 Monte-Carlo simulations.}
	\label{fig:estimates}
\end{figure}

Geant4 \cite{geant} simulations were used to estimate what to expect from SABRE in terms of background and sensitivity. It was assumed that the crystals will have the same purity as found in the Astrgrade powder and external radiation can be shielded appropriately. The resulting background spectrum is shown in Figure~\ref{fig:est-back}. The blue line shows the total background without vetoing any events. Rejecting all events that either have multiple hits inside the crystals or are accompanied by a hit inside the veto results in the background spectrum shown in red with an average level of 0.15~cts/(keV kg d). The figure also shows that the majority of the background comes from the crystals (shown in green). SABRE is therefore working to further reduce the radioactive contamination inside the final crystals compared to the currently available powder.

Assuming that DAMA's modulation is indeed due to WIMPs and no other seasonal effects are present in the region of interest, SABRE's sensitivity was estimated. Using the above background prediction and a 50-kg (25-kg) target, SABRE has a $4\sigma$ ($2.5\sigma$) power to confirm DAMA's results in a 3-year run, see Figure~\ref{fig:est-sens}. Similar sensitivity is found to validate that the modulation is not due to dark matter.

\section{Current Status}

\subsection*{Proof of Principle}
SABRE is currently in a proof-of-principle phase with two primary goals. Firstly the impurity levels of the first high-purity crystal will be measured by gamma counting. This is especially important for isotopes where the ICP-MS is not sensitive enough. $^{40}$K might be such a case since the ICP-MS currently has a detection limit of 2~ppb. Secondly the effectiveness of the veto, so far only estimated by simulations, shall be validated by measurements. The proof-of-principle phase will take place at LNGS in Italy where parts of the former Icarus site in Hall B were approved for this purpose. Temporary shielding made out of lead, water tanks, and maybe polyethylene will be used. Most components are already available, commissioning of the veto vessel started at Princeton University, and preparations at LNGS are in progress.

\subsection*{Full Scale Experiment}
For the full-scale experiment, twin detectors are currently planned. The first detector will be placed at LNGS, Italy, the second one in the Stawell Underground Physics Laboratory (SUPL) in Australia. This setup allows us to take measurements in the northern and southern hemisphere and thus verify if other seasonal effects or the site have an influence on the measurement, thus reducing systematic effects. The proof-of-principle experiment at LNGS will be upgraded to the full-scale setup. The funding for the laboratory in Australia is secured as well as part of the funding for the experiment and the final design in progress.

\subsection*{Stawell Underground Physics Laboratory}
SUPL will be the first underground laboratory in the southern hemisphere and is located $\sim240$~km west of Melbourne. A part of a working gold mine will be converted into the lab and the chosen site is $\sim1.02$~km deep with a flat overburden, corresponding to $\sim3$~km water equivalent (similar to LNGS). Electricity and optical fibers are already available and the lab can be reached by car or truck. Construction is planned for 2016 and it is foreseen that detectors can start to be set up beginning of 2017.

\subsection*{Collaboration}
The current SABRE collaboration consists of around 30 members primarily from Princeton University (USA), INFN (Italy), and different universities in Australia; no member of the DAMA collaboration is directly involved in SABRE.

\section{Conclusion}
Interpreting DAMA's modulations as a WIMP signal is in tension with the results of other experiments. Since those experiments use a different target, an independent experiment using NaI(Tl) crystals like DAMA is needed and planned with the SABRE experiment. The necessary ultra-high-purity crystals are currently in preparation. A lower than DAMA background is planned to be achieved by using high-purity materials and an active veto. A higher light yield and lower threshold is expected due to improved, pre-amplified PMTs, which are directly coupled to the crystal.
A proof-of-principle is currently in preparation at LNGS. Preparations for two full-scale twin detectors started, which will be located at LNGS (Italy) and SUPL (Australia).

\medskip
\ack{The SABRE NaI(Tl) program is supported by NSF Grants PHY-0957083, PHY-1103987, PHY-1242625, and PHY-1506397. SABRE is also supported by INFN, LNGS as well as ARC LIEF Grant LE160100080, ARC LP150100705, and the National Stronger Regions Fund and Regional Jobs and Infrastructure Fund.}
\bigskip

\bibliography{bibliography}

\providecommand{\newblock}{}
\begin{thebibliography}{10}
\expandafter\ifx\csname url\endcsname\relax
  \def\url#1{{\tt #1}}\fi
\expandafter\ifx\csname urlprefix\endcsname\relax\def\urlprefix{URL }\fi
\providecommand{\eprint}[2][]{\url{#2}}

\bibitem{dama2013}
{Bernabei} R {\em et~al.\/} (DAMA/LIBRA Collaboration) 2013 {\em European
  Physical Journal C\/} {\bf 73} 2648 (\textit{Preprint} \eprint{1308.5109})

\bibitem{lux}
Akerib D {\em et~al.\/} (LUX Collaboration) 2014 {\em Phys.Rev.Lett.\/} {\bf
  112} 091303 (\textit{Preprint} \eprint{1310.8214})

\bibitem{xenon100}
Aprile E {\em et~al.\/} (XENON100) 2012 {\em Phys. Rev. Lett.\/} {\bf 109}
  181301 (\textit{Preprint} \eprint{1207.5988})

\bibitem{supercdms}
Agnese R {\em et~al.\/} (SuperCDMS Collaboration) 2014 {\em Phys.Rev.Lett.\/}
  {\bf 112} 241302 (\textit{Preprint} \eprint{1402.7137})

\bibitem{sabre2015}
Xu J, Calaprice F, Froborg F, Shields E and Suerfu B 2015 {\em AIP Conf.
  Proc.\/} {\bf 1672} 040001
  \urlprefix\url{http://scitation.aip.org/content/aip/proceeding/aipcp/10.1063/1.4927983}

\bibitem{Gluscevic2015}
Gluscevic V, Gresham M~I, McDermott S~D, Peter A~H~G and Zurek K~M 2015 {\em
  JCAP\/} {\bf 1512} 057 (\textit{Preprint} \eprint{1506.04454})

\bibitem{Lee2013}
Lee S~K, Lisanti M, Peter A~H~G and Safdi B~R 2014 {\em Phys. Rev. Lett.\/}
  {\bf 112} 011301 (\textit{Preprint} \eprint{1308.1953})

\bibitem{damaapp}
{Bernabei} R {\em et~al.\/} (DAMA/LIBRA Collaboration) 2008 {\em Nucl. Instrum.
  Methods A\/} {\bf 592} 297 -- 315 ISSN 0168-9002
  \urlprefix\url{http://www.sciencedirect.com/science/article/pii/S016890020800675X}

\bibitem{Bridgman1925}
Bridgman P~W 1925 {\em Proceedings of the American Academy of Arts and
  Sciences\/} {\bf 60} 305--383 ISSN 01999818
  \urlprefix\url{http://www.jstor.org/stable/25130058}

\bibitem{pmtback_2015}
Aprile E {\em et~al.\/} (XENON) 2015 {\em Eur. Phys. J.\/} {\bf C75} 546
  (\textit{Preprint} \eprint{1503.07698})

\bibitem{EmilyThesis}
Shields E 2015 {\em {SABRE: A search for dark matter and a test of the
  DAMA/LIBRA annual-modulation result using thallium-doped sodium-iodide
  scintillation detectors}\/} Ph.D. thesis Princeton University

\bibitem{geant}
Agostinelli S {\em et~al.\/} 2003 {\em Nucl. Instrum. Methods A\/} {\bf 506}
  250 -- 303 ISSN 0168-9002
  \urlprefix\url{http://www.sciencedirect.com/science/article/pii/S0168900203013688}

\end{thebibliography}
\bibliographystyle{iopart-num}

\end{document}